\documentclass[aps,prl,twocolumn,showpacs,subfigure,superscriptaddress,nobibnotes,nofootinbib]{revtex4}
\usepackage[dvips]{graphicx}
\usepackage{epsfig}
\usepackage{bm}   % for adding bold symbols etc. in maths mode
\usepackage{dcolumn}% Align table columns on decimal point
\usepackage{graphicx}
\usepackage{rotate}
\usepackage{amsfonts}

\begin{document}

\title{Identification of mixed-symmetry states in an odd-mass nearly spherical nucleus\\}

\author{J. N.~Orce}
\email{jnorce@pa.uky.edu}
\homepage{http://www.pa.uky.edu/~jnorce}
\affiliation{Department of Physics and Astronomy, University of Kentucky, Lexington,
 Kentucky 40506-0055, USA}

\author{J. D. Holt}
\affiliation{Nuclear Structure Laboratory, Department of Physics and Astronomy, 
SUNY, Stony Brook, NY 11794-3800, USA}

\author{A. Linnemann}
\affiliation{Institut f\"ur Kernphysik, Universit\"at zu K\"oln, 50937 K\"oln, Germany}

\author{C. J. McKay}
\affiliation{Department of Physics and Astronomy, University of Kentucky, Lexington, Kentucky 40506-0055, USA}

\author{S. R. Lesher}
\affiliation{Department of Physics and Astronomy, University of Kentucky, Lexington, Kentucky 40506-0055, USA}

\author{C. Fransen}
\affiliation{Institut f\"ur Kernphysik, Universit\"at zu K\"oln, 50937 K\"oln, Germany}

\author{J. W. Holt}
\affiliation{Nuclear Structure Laboratory, Department of Physics and Astronomy, SUNY, Stony Brook, NY 11794-3800, USA}

\author{A. Kumar}
\affiliation{Department of Physics and Astronomy, University of Kentucky, Lexington, Kentucky 40506-0055, USA}

\author{N. Warr}
\affiliation{Institut f\"ur Kernphysik, Universit\"at zu K\"oln, 50937 K\"oln, Germany}

\author{V. Werner}
\affiliation{Institut f\"ur Kernphysik, Universit\"at zu K\"oln, 50937 K\"oln, Germany}

\author{J. Jolie}
\affiliation{Institut f\"ur Kernphysik, Universit\"at zu K\"oln, 50937 K\"oln, Germany}

\author{T. T. S. Kuo}
\affiliation{Nuclear Structure Laboratory, Department of Physics and Astronomy, SUNY, Stony Brook, NY 11794-3800, USA}

\author{M. T. McEllistrem}
\affiliation{Department of Physics and  Astronomy, University of Kentucky, Lexington, Kentucky 40506-0055, USA}

\author{N. Pietralla}
\affiliation{Nuclear Structure Laboratory, Department of Physics and Astronomy, SUNY, Stony Brook, NY 11794-3800, USA}
\affiliation{Institut f\"ur Kernphysik, Universit\"at zu K\"oln, 50937 K\"oln, Germany}

\author{S. W. Yates}
\affiliation{Department of Physics and Astronomy, University of Kentucky, Lexington, Kentucky 40506-0055, USA}
\affiliation{Department of Chemistry, University of Kentucky, Lexington, Kentucky 40506-0055, USA}

\date{\today}


\date{\today}

\begin{abstract}

The low-spin structure of $^{93}$Nb has been studied using the (n,n$'$$\gamma$) reaction
at neutron energies ranging from 1.5 to 3 MeV and the $^{94}$Zr(p,2n$\gamma$)$^{93}$Nb 
reaction at bombarding energies from 11.5 to 19 MeV.
%Excitation functions, lifetimes, and branching ratios were measured,
%and multipolarities and spin assignments were determined.
States at  1779.7 and 1840.6 keV, respectively, are proposed as
mixed-symmetry states associated with the $\pi2{p^{-1}_{1/2}}\otimes(2_{1, {\rm MS}}^{+},^{94}$Mo) 
coupling. These assignments are derived from the observed $M1$ and $E2$ transition strengths 
to the 2${p_{1/2}^{-1}}$
symmetric one-phonon states, energy systematics, spins and parities, and
comparison with shell model calculations.
\end{abstract}


\maketitle
%

%Albeit simple, the Interacting Boson Model-2 (IBM-2) provides suitable predictions of
%nuclear collective excitations in  heavy even-mass nuclei \cite{1}.
%This group theoretical or algebraic model truncates the shell-model space reducing dramatically
%the basis states by introducing symmetry requirements on boson structures.
%Fermions of integer spin are finally treated as bosons.
%In the IBM-2, the isospin formalism can be extended to proton and neutron bosons, introducing
%F-spin,  the boson quantum number which characterizes the proton-neutron symmetry ($pn$) of
%nuclear states.
%By coupling the proton and neutron degrees of freedom in a nucleus with N$_{\pi}$
%proton pairs and N$_{\nu}$ neutron pairs, F-spin assumes values from
%F$_{max}$=(N$_{\pi}$ $+$ N$_{\nu}$)/2 to F$_{min}$=(N$_{\pi}$ $-$ N$_{\nu}$)/2.
%Fully symmetric or multiphonon states correspond then to F=F$_{max}$. However, IBM-2 wavefunctions
%with F$\ <$F$_{max}$ possess at least one pair of proton or neutron bosons
%which are not symmetric.
%$|\langle J{^{f}_{sym}},F_{max}|M1|J{^{i}_{\rm MS}},F_{max}-1\rangle| \approx$

The interplay of collective and single-particle excitations in nuclei provides an
excellent testing ground to examine the coupling between bosonic and fermionic 
degrees of freedom. 
%Quadrupole and octupole vibrations in  nuclei have been well characterized; 
%however, information about new types of basic nuclear excitations is emerging.
%The Interacting Boson Model (IBM) originally proposed that collective states of nuclei
%could be characterized as composed of bosons formed from coupled proton and neutron pairs.
%Nuclei could be well described as boson representations of subgroups of the U(6) group.
%To match actual nuclear excitations, separate group representations were needed for proton
%and neutron bosons, leading to IBM-2 \cite{1}.  This separation of proton and neutron bosons led
%directly to a new class of collective excitation so-called  mixed-symmetry (MS) states.
In the interacting boson model with separate representations for proton and neutron bosons (IBM-2) \cite{1},
a new class of collective excitations,  mixed-symmetry (MS) states, arises.
MS states are collective vibrational phenomena which are not
fully symmetric with respect to the proton-neutron ($pn$)  degree of freedom \cite{2}, 
presenting $pn$ symmetry departures from ground-state symmetry. 
These excitations are of isovector character, i.e., proton and neutron spin contributions
are additive in the vector part of the $M1$ magnetic dipole operator and may lead to strong
$M1$ transitions (matrix elements of about 1$\mu_N$) from the $MS$ state. 
% generated by the operator
%(s$_\nu^\dagger$d$_\pi^\dagger$ - s$_\pi^\dagger$d$_\nu^\dagger$),
%to a fully symmetric state,  (s$_\nu^\dagger$d$_\pi^\dagger$ + s$_\pi^\dagger$d$_{\nu}^\dagger$).
A systematic description of  these  collective nuclear states within the  IBM-2   can be
found in Ref.~\cite{isa86}.
Recently, $MS$ states have also been examined from a shell model approach \cite{3}.

Whereas $MS$ states have been studied extensively in even-even nuclei, little
is known about 
$MS$ states in odd-mass nuclei, as only the  scissors mode has
been identified in a few deformed nuclei through strong dipole
transitions to the ground-state \cite{firstscissors}. The nearly-spherical
N=52 isotones form a  bountiful region for $MS$ findings, where
2$_{1,MS}^+$ states (of particular interest to this work) and 
[$2_1^+\otimes2_{1,MS}^+$] two-phonon $MS$ states  have been identified
\cite{norbert,cfransen}. The characterization of such states
indicates that the  $2^+_{1, {\rm MS}}$ state acts
as a building block of  nuclear excitations.
In $^{92}_{40}$Zr, the proton subshell closure at Z$=$40 results in a much stronger individual particle
strength than in the isotone $^{94}_{42}$Mo.
However, a somewhat collective $MS$ state  was identified as the $2_2^+$ level at
1.847\,MeV. This $2_2^+$ state  exhibits  a large \linebreak
B(M1; 2$_{1, {\rm MS}}^{+}$$\rightarrow$2$_{1}^{+}$)=0.37(4)
$\mu_N^{2}$ and a weakly collective B(E2; 2$_{1, {\rm MS}}^{+}$$\rightarrow$0$_{1}^{+}$)=3.4(4) W.u. \cite{fransen92zr}.
The nucleus $^{94}_{42}$Mo  exhibits  more collective features and the $2^+_{\rm 1,MS}$ state was clearly
identified as the $2_3^+$ state at 2.067\,MeV [6]. It  displays a stronger $M1$ transition,
B(M1; 2$_{1, {\rm MS}}^{+}$$\rightarrow$2$_{1}^{+}$)=0.56(5) $\mu_N^{2}$ and
a weakly collective $E2$ transition to the ground-state,
B(E2; 2$_{1, {\rm MS}}^{+}$$\rightarrow$0$_{1}^{+}$)=2.2(2) W.u. \cite{fransen}. In a weak-coupling scenario, $MS$ states in the odd-Z N=52 isotone, $^{93}_{41}$Nb,
might be expected at similar energies ($\sim$2 MeV)  as their
even-Z, N=52 isotone neighbors. If found in an odd-mass nucleus, these $MS$ states 
would give evidence for weak-coupling of the fermion to the collective excitations and 
affirm the effectiveness of IBM-2 in separating proton and neutron representations.  
%There is no reason that the group symmetries, which inspired the original IBM concept,
%should not apply to odd-mass nuclei as well as even-mass.

Excited states in $^{93}$Nb  can be regarded as  resulting from the coupling
of a  1$g_{9/2}$ proton to a $^{92}_{40}$Zr core, and a  2$p_{1/2}^{-1}$ proton-hole to a $^{94}_{42}$Mo core
\cite{93nb_old,93nb_coulomb}. These couplings  result in two independent and unmixed
phonon structures of opposite
parity. A quintet of positive-parity
states  built on the J$^\pi$=9/2$^+$ ground-state results from the
$\pi 1g_{9/2}$$\otimes$(2${^+_1}$,$^{92}$Zr) particle-core coupling. This quintet has been identified
in agreement with expectations of the center-of-gravity theorem \cite{centerofgravity,93nb_old}, and
the assignment is supported by Coulomb excitation results  \cite{kregar}. A simpler structure,  a doublet
of negative-parity states
built on the J$^\pi$=1/2$^-$ first excited state at 31 keV (with a half-life of 16 years),
is also observed and corresponds to the $\pi2p_{1/2}^{-1}$$\otimes$(2${^+_1}$,$^{94}$Mo)
configuration. 
%There are few realistic models of $MS$ states in these nuclides, due probably to the increasing complexity
%arising from the interplay of $MS$, phonon, intruder and single-particle degrees of freedom,
%which lead to admixtures in the  nuclear wavefunctions \cite{heyde2}.
%Experimentally, angular momenta  carried by the unpaired particle
%and further particle-core couplings lead to several phonon states lying at similar excitation energies
%as those expected for the $MS$ states. A distribution of phonon and $MS$ strength among
%several levels is expected \cite{heyde2,nord2}. Symmetric phonon excitations, however,
%present  different decay signatures than that of a $MS$ state, and  decay preferentially through collective electric quadrupole
%transitions following the $\Delta n_\lambda = \pm 1$
%selection rule for vibrational states (where $n_\lambda$ is the phonon number).
We have identified $MS$ states in the negative-parity structure of the nearly spherical
odd-Z nucleus $^{93}$Nb, which correspond to the 2$_{1,MS}^+$  states found in neighboring
even-even nuclei. Identifications are based on $M1$ and $E2$ strengths, energy systematics,
and spin-parity assignments and from the comparison with shell model calculations using
the low-momentum nucleon-nucleon interaction, $V_{low-k}$ \cite{bogner03}.

\begin{figure}[h]
\begin{center}
\includegraphics[width=3.4cm,height=5.6cm,angle=-90]{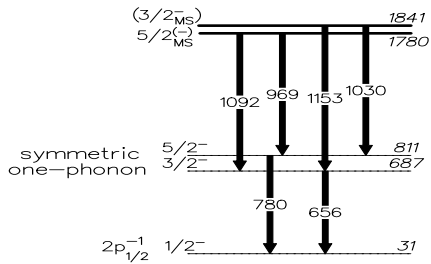}
\caption{$^{93}$Nb partial level scheme showing the 2p$_{1/2}^{-1}$ single-particle, symmetric one-phonon
and proposed $MS$  states of importance to this discussion.}
\label{fig:scheme}       % Give a unique label
\end{center}\vspace{-5mm}
\end{figure}

The nucleus $^{93}$Nb was studied using the (n,n$^\prime$$\gamma$) reaction  at
the University of Kentucky and the   $^{94}$Zr(p,2n$\gamma$$\gamma$)$^{93}$Nb reaction at
the University of Cologne. Excitation functions, lifetimes and 
branching ratios were measured using the $^{93}$Nb(n,n$^\prime$$\gamma$)
reaction \cite{nng} at neutron energies ranging from 1.5 to 2.6 MeV. 
%Neutrons were provided by the 7 MV electrostatic accelerator at
%the University of Kentucky through the $^3$H(p,n)$^3$He reaction. The scattering sample was 56g (3$\times$2 cm cylindrical)
%of natural $^{93}$Nb (only stable isotope).
%Pulsed-beam techniques were used to reduce background, with beam-pulses
%separated by 533 ns and bunched to about 1.0 ns.
%This allowed  the use of
%the time-of-flight technique for background suppression by gating on the
%appropriate prompt time windows \cite{pulsed}. Finally, the collected $\gamma$ rays
%were detected using a BGO Compton-suppressed 55\% HPGe spectrometer with 2.0 keV resolution.
%Both excitation functions and angular
%distributions were normalized to the neutron flux.
Gamma-gamma  coincidences  were also measured with
the (n,n$'$$\gamma$$\gamma$) reaction at a neutron energy of 3 MeV.
Excitation functions, together with the analysis of background substracted coincidence spectra 
allowed the construction of a comprehensive level scheme up to 2.1 MeV.
% in order to identify
%the decay pathways, measure their branching ratios, and confirm the decay scheme from
%excitation functions. 
The coincidence methods following neutron inelastic scattering are described 
by McGrath {\it et al.} \cite{mcgrath}. Lifetimes were measured through the Doppler-shift
attenuation method following the (n,n$'$$\gamma$) reaction \cite{belgya}.
%The shifted
%$\gamma$ ray energy is given by $E_{\gamma}(\theta_\gamma)=E_{\gamma_{0}} [1+\frac{v_{0}}{c}F(\tau)cos{\theta_\gamma}]$,
%with $E_{\gamma_{0}}$ being the unshifted  $\gamma$ ray
%energy, $v_{0}$ the initial maximum recoil velocity in the center of mass frame, $\theta$ the angle of
%observation and $F(\tau)$ the attenuation factor, which is related to the
%nuclear stopping process described by Blaugrund \cite{blaugrund}. Finally, the lifetimes of the states
%can be determined by comparison with the $F(\tau)$ values calculated using the Winterbon formalism
%\cite{winterbon}.
From an angular correlation experiment with the $^{94}$Zr(p,2n$\gamma$$\gamma$)$^{93}$Nb
reaction, branching ratios were measured and multipolarities and  spin assignments 
determined. 
%The proton beam was provided by the 10 MV TANDEM accelerator at the University
%of Cologne, bombarding a 2mg/cm$^2$ thick $^{94}$Zr target enriched to 96.93$\%$,
%with a beam current ranging from 2 to 3 $\mu$A and beam energies ranging from
%11.5 to 19 MeV. The de-excited $\gamma$ rays were detected using the HORUS spectrometer,
%comprised of 16 HPGe detectors with the following features: a) six of these formed the
%EUROBALL cluster detector \cite{euroball} with the central detector placed at
%$\theta$=90$^{\circ}$ with respect to the beam axis and $\phi$=0$^{\circ}$, where
%$\phi$ is the clockwise angle around the beam axis; b) four with a relative efficiency of
%55\% and placed at $\theta$=45$^{\circ}$ and 135$^{\circ}$ in a vertical plane above
%and below the beam axis were complemented by Compton
%suppression shields;
%c) the remaining detectors with a relative efficiency of 25-35 \% were placed in the
%$\theta$=90$^{\circ}$ plane and at angles of $\phi$=55$^{\circ}$, 125$^{\circ}$, 235$^{\circ}$
%and 305$^{\circ}$.
%Given the possible detector combinations, this detector arrangement has nine possible angular correlation groups.
%Due to problems concerning the efficiency of the cluster
%detector, only 7 out of the 9 possible  correlation groups were used during this analysis.
We used seven angular correlation groups defined by three angles: $\theta_1$ and $\theta_2$ are the angles of the two detector
axes relative to the beam axis, and the angle $\phi$ between the planes defined by the beam axis and
emission directions of the $\gamma$-rays.
When fitting angular correlation data, we fixed the Gaussian width, $\sigma$, of the m-substate
distribution from fits to previously known transitions in $^{93}$Nb, and treated 
the multipole mixing ratio, $\delta$ \cite{delta},  of the unknown transition as a free parameter  
following the formalism
developed by  Krane, Steffen and Wheeler \cite{angcorr}.
%Consistent results are obtained by gating on a $\gamma$ ray transition either above or below the transition of interest.
The current results for the strongest populated
transitions are  in good agreement  with those identified in previous work \cite{93nb_old,93nb_coulomb}.
Table \ref{tab:strengths}
gives the properties of  selected negative-parity levels in $^{93}$Nb.
The proposed $MS$ states  are the 1779.7 keV and 1840.6 keV levels
(as shown in Figure \ref{fig:scheme}) arising from the
$2{p_{1/2}^{-1}}\otimes(2_{1, {\rm MS}}^{+}, ^{94}$Mo) configuration.

%\vspace{0mm}
%\begin{figure}[]
%\begin{center}
%\includegraphics[width=7.4cm,height=7.3cm,angle=-0]{angcorr.eps}
%\caption{Angular correlation plots for a) the 385 and 959 keV $\gamma$ ray transitions
%(17/2$^+$ $\rightarrow$ 13/2$^+$ $\rightarrow$ 9/2$^+$)  (top)  giving a mixing ratio, $\delta$, of
%0.024(33) for the 385 keV $\gamma$ ray, and b) the 338 and 744 keV transitions  (9/2$^+$ $\rightarrow$ 7/2$^+$
%$\rightarrow$ 9/2$^+$)  (bottom) with $\delta$$=-$0.087(23) for the 338 keV $\gamma$ ray.
%These results are in good agreement with previous work, where $\delta$$=$0 and -0.12(5)
%were determined for the 385 and 338 keV transitions, respectively.
%The fitting parameter $\sigma$ is the width of the magnetic substate distribution found
%to be between about 2 and 4. }
%\end{center}
%\end{figure}

%\begin{figure*}[]\includegraphics[width=5.4cm,height=15.3cm,angle=-90]{msregion.eps}\caption{}\label{fig:level}\end{figure*}

%The analysis of the data indicates two well defined 
%and nearly separated
%structures as the main characteristic of this nucleus. One is built on the ground state and positive parity
%one-phonon states and similarly, the other structure is built on the $1/2^-$ excited state and negative parity one-phonon
%states. We have search for $MS$ states in the negative-parity side due to a simpler decay scheme.

\vspace{0mm}
\begin{figure}[]
\begin{center}
\includegraphics[width=6.5cm,height=7cm,angle=-0]{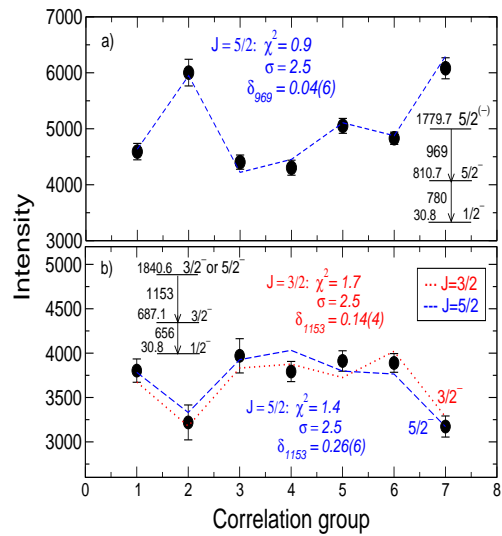}
\caption{(Color online) Angular correlation data for a) the 1092 keV transition gated by the 656 keV
$\gamma$-ray 
%(5/2$^-$ $\rightarrow$ 3/2$^-$ $\rightarrow$ 1/2$^-$)
and b) the 1153 keV transition gated by the 656 keV $\gamma$-ray. Spin and
parity assignments of $J^\pi=3/2^-$ and $5/2^-$ are equally possible for the 1840.6 keV level
with fits to the data of $\delta_{1153}=0.14(4)$ and $0.26(6)$, respectively. 
The Gaussian width of the magnetic substate
distribution, $\sigma=$2.5, was fixed from fits to previously known
transitions in $^{93}$Nb. The $\chi^2$ values given in our angular correlation fits 
are reduced, and are given by the total $\chi^2$ divided by the``number of correlation groups $-$ 2''.}
%Dashed lines indicate other angular correlation fits with unlikely
%spins for the  1297.4 keV level.
%The fitting parameter $\sigma$ is the width of the magnetic substate distribution, usually  found
%to be between  2 and 3. }
\label{fig:1}       % Give a unique label
\end{center}
\end{figure}

\begin{figure}[]
\vspace{-3mm}\begin{center}
\includegraphics[width=6.cm,height=5.cm,angle=-0]{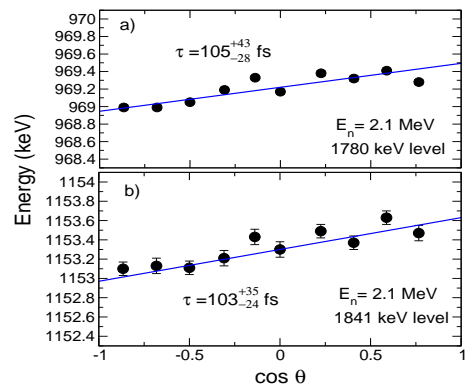}
\caption{(Color online) Doppler-shift attenuation data for $\gamma$-rays de-exciting a) the 1779.7 keV level
and b) the 1840.6 keV level.}
\label{fig:3}       % Give a unique label
\end{center}\vspace{-8mm}
\end{figure}

\begin{table*}[!]
\begin{center}
\caption{
Results obtained from the $^{93}$Nb(n,n$^{\prime}$$\gamma$) and $^{94}$Zr(p,2n)$^{93}$Nb
measurements. Negative-parity levels, lifetimes, initial and final spin of
the states,  $\gamma$-ray energies, branching and mixing ratios, and experimental
B(M1)$\downarrow$ and B(E2)$\downarrow$ transition rates are listed.
Shell model B(M1$\downarrow$) and B(E2$\downarrow$) predictions for relevant transitions together with
isoscalar (IS) and isovector (IV) components of the $E2$ operator are shown in the
last four columns.
An asterisk labels newly identified levels and $\gamma$-ray transitions.}
\label{tab:strengths}
\begin{tabular}{lcccccccccccc}
\hline \hline \\
$E_L$       & \hspace{3mm} $\tau$  & &  $J^\pi_i$ $\rightarrow$ $J^\pi_f$ & \hspace{3mm} $E_\gamma$  & $I_\gamma$ & \hspace{3mm} $\delta$  & \hspace{3mm} B($M1$)  	& \hspace{3mm} B($E2$)    & \hspace{3mm} $B(M1)$ & $B(E2)$ & IS & IV \\

   (keV)    & \hspace{3mm} (fs)    & &                                    & \hspace{3mm} (keV)       &            & \hspace{3mm}           & \hspace{3mm} $(\mu_N^2)$   & \hspace{3mm}  (W.u.)    &  \multicolumn{2}{c}{\hspace{3mm}[Shell Model]} && \\ \hline

%1284.8(2)      &  $252^{+81}_{-51}$    &  &  $\frac{5}{2}^-$ \hspace{3mm}    $\frac{1}{2}^-$  &  1253.5          &  100(4)  &   $E2$       &                         & $32^{+10}_{-9}$    &- &-&-&- \\ [0.7ex]
%
%               &              &  &                           \hspace{8.7mm}  $\frac{3}{2}^-$  &   597.3$^*$      &   25(4)  &  $0.14(4)$   & $0.20^{+0.10}_{-0.08}$  & $6^{+3}_{-2}$      & -&-&-&- \\[0.7ex]
%
%               &              &  &                           \hspace{8.7mm}  $\frac{5}{2}^-$  &   473.9$^*$      &    5(4)  &   $(M1)$     &    $\ <$0.08            &                    &  -&-&-&- \\[0.7ex]

%1370.1(2)      &   $\ > 789$  &  &  $\frac{5}{2}^-$ \hspace{3mm}    $\frac{3}{2}^-$  &  683.2$^*$   &   30(4)  &  $-0.34(5)$     &  $\ <$0.05           &  $\ <$7         & -&-&-&-  \\ [0.7ex]

%               &              &  &                  \hspace{8.7mm}  $\frac{5}{2}^-$  &  559.4       &  100(4)  &  $-0.32(7)$     &  $\ <$0.29            &  $\ <$54&  -&-&-&- \\ [0.7ex]

1395.8(2)      &   $\ > 792$  &  &  $\frac{7}{2}^-$ \hspace{3mm}    $\frac{3}{2}^-$  &   708.6      &   9(4)  & $E2$             &                  &    $\ <$18            &- & 8.50 & 28.61 & -13.85  \\[0.7ex]

               &              &  &                  \hspace{8.7mm}  $\frac{5}{2}^-$  &   585.1      & 100(4)  & $-0.10(2)$       &    $\ <$0.31     &   $\ <$5.2                & 0.00003 & 0.879 & 9.17  & -4.39  \\[0.7ex]

1500.0(1)     &  1164(301)  &  &  $\frac{9}{2}^-$  \hspace{3mm}  $\frac{5}{2}^-$    &   689.5     &  18(3)      & $E2$  &             &   $27^{+15}_{-9}$    &  -  & 10.63 & 35.39  & -15.21  \\[0.7ex]

1572.1(2)      &   $281^{+213}_{-94}$    &  &  $\frac{3}{2}^-$ \hspace{3mm}    $\frac{3}{2}^-$  &  885.1$^*$  &   37(5)   &    $-1.60(14)$      &   $0.02(1)$                 &  $36^{+27}_{-19}$  &0.002 & 14.46 & 25.71  & -8.85  \\ [0.7ex]

               &                         &  &                  \hspace{8.7mm}  $\frac{5}{2}^-$  &  761.4$^*$  &  100(5)   &    $-0.28(3)$       &   $0.27^{+0.16}_{-0.12}$    &  $21^{+12}_{-10}$ & 0.00001 & 6.0 & 16.66 & -6.44 \\[0.7ex]

               &                         &  &                  \hspace{8.7mm}  $\frac{5}{2}^-$  &  287.4$^*$  &   20(5)   &      $(M1)$         &   $\ <$1.09                          &                     \\[0.7ex]

1588.4(2)$^*$  &   $\ > 1260$            &  &  $\frac{5}{2}^-$ \hspace{3mm}    $\frac{3}{2}^-$  &  901.2$^*$  & 100(8)   &    $-0.53(6)$                  &  $\ <$0.04     &     $\ <$8  & 0.0012 & 4.36 & 17.18& -5.26  \\ [0.7ex]

               &                         &  &                  \hspace{8.7mm}  $\frac{5}{2}^-$  &  777.8$^*$  &  18(8)   &    $-4.03^{+1.32}_{-3.45}$    &  $\ <$0.001     &     $\ <$13 & 0.0069 &16.94 &34.19 & -12.37 \\[0.7ex]

1779.7(2)      &   $105^{+43}_{-28}$  &  & \hspace{-1.2mm}  $\frac{5}{2}^{(-)}$ \hspace{1mm}    $\frac{3}{2}^-$  &  1092.4$^*$  & 8(5)   & $0.05(9)$     &  $0.03^{+0.04}_{-0.02}$     &  $0.04^{+0.04}_{-0.03}$ &0.037 & 0.306 &-0.74 & 2.35  \\ [0.7ex]

               &                      &  &                  \hspace{8.7mm}  $\frac{5}{2}^-$  &   969.0      & 100(5) & $0.04(6)$     &  $0.55^{+0.24}_{-0.18}$     &  $0.5(2)$ & 0.616 & 0.0001 &-0.72 & 4.92 \\[0.7ex]

               &                      &  &                  \hspace{8.7mm}  $\frac{1}{2}^-_{gs}$  &   & & & & & - & 4.28 & 12.63 & 19.87 \\[0.7ex]

1840.6(2)$^*$  &   $103^{+35}_{-24}$   &  & \hspace{-4mm}  $\frac{3}{2}^{-}, \frac{5}{2}^-$ \hspace{0mm}    $\frac{3}{2}^-$  &  1153.4$^*$  & 100(4)  & $0.14(4),0.26(6)$     &  $0.29(8),0.28(9)$     &  $2.5(7),8(2)$ & 0.462 & 0.306 & -4.36  & 4.90 \\

               &                       &  &                                                 \hspace{8.7mm}  $\frac{5}{2}^-$  &  1029.6$^*$  & 20(4)   & $0.32(6),0.17(8)$     &  $0.08(3),0.08(3)$     &  $4(1),1(1)$ & 0.046 & 0.078 & -1.91 & 2.63 \\[0.7ex]

 & & & \hspace{8.7mm}  $\frac{1}{2}^-_{gs}$ & &&&& & - & 5.99 & 13.03 & 14.55 \\[0.7ex]

1948.1(2)      &  $230^{+133}_{-69}$  &  & \hspace{-1.2mm}  $\frac{7}{2}^{(-)}$ \hspace{1mm}    $\frac{5}{2}^-$  &  1137.4      & 100    & $0.05(4)$  &  $0.17(7)$     &  $0.2(1)$\\[0.7ex]

1997.6(2)$^*$  &   $92^{+22}_{-17}$   &  &  $\frac{5}{2}^-$ \hspace{3mm}    $\frac{3}{2}^-$  &  1310.2$^*$  & 12(5)  & $-0.29(12)$   &  $0.03(2)$     &  $0.8^{+0.6}_{-0.4}$\\[0.7ex]

               &                      &  &                  \hspace{8.7mm}  $\frac{5}{2}^-$  &  1186.9$^*$  & 100(5) & $-0.31(11)$   &  $0.30^{+0.09}_{-0.07}$     &  $12^{+4}_{-3}$\\[0.7ex]

%2012.6(2)$^*$  &                      &  &  $\frac{3}{2}^-$  \hspace{3mm}    $\frac{3}{2}^-$  &  1325.8$^*$  & 100(5)    & $4.47^{+1.53}_{-0.94}$ &     &  \\

%               &                      &  &                   \hspace{8.7mm}  $\frac{3}{2}^-$  &  440.4$^*$   &   6(5)    &                        &     &  \\[0.7ex]

% 2024.4(2)$^*$  &  $78^{+41}_{-25}$    &  & $\frac{3}{2}^-$   \hspace{3mm}    $\frac{3}{2}^-$  &  1337.1$^*$  &  100(3)    & $-4.70^{+0.84}_{-1.28}$   & $0.01(1)$      &  $91^{+76}_{-53}$    \\[0.7ex]
%
%               &                      &  &                   \hspace{8.7mm}  $\frac{3}{2}^-$  &  452.1$^*$   &   3(3)     &     $(M1)$                & $\ <0.23$      &  \\[0.7ex]

%2099.6(2)     &  $133^{+62}_{-36}$    &  & \hspace{-1.4mm}  $\frac{7}{2}$  \hspace{5.3mm}   $\frac{5}{2}^-$  &  1288.9$^*$ &  46(7)    & $-0.05(5)$    &  $0.05(1)$     & $0.2^{+0.3}_{-0.1}$      \\
%
%              &                       &  &                       \hspace{8.7mm} $\frac{7}{2}^-$  &   703.8     & 100(7)    &               &                &                          \\[0.7ex]

 \hline \hline
 \vspace{-1.2cm}
\end{tabular}
\end{center}
\end{table*}

%\begin{figure*}[]
%\begin{center}
%\includegraphics[width=7cm,height=18.8cm,angle=-90]{nega.eps}
%\caption{$^{93}$Nb partial level scheme showing the positive and negative parity one-phonon states
%(left side), and including the negative-parity states with their respective branches.
%The thickness of the arrows is proportional
%to the competing branches (relative intensities) depopulating a particular state, which were measured
%from the $^{94}$Zr(p,2n$\gamma\gamma$) coincidence data.}
%\label{fig:scheme}       % Give a unique label
%\end{center}
%\end{figure*}
%Figure \ref{fig:2} shows the partial level scheme of interest for $^{93}$Nb and 

{\it 1779.7 keV state} ---  The previously proposed (5/2$^-$) level at 1779.7 keV yields
%{\bf (Note: here we really have several possible fragments of the 5/2$^-$ $MS$ state (1284, 1779.7, 1840.6
%and 1997 keV levels) and only shell model calculations could give a further insight)},
a  new 1092 keV branch to the first 3/2$^-_1$ excited state that
has been revealed  from the excitation function and coincidence data. The level has been
assigned as  J$^{\pi}$=5/2$^{(-)}$ by the analysis
of the angular correlation data (see Table \ref{tab:strengths} and top panel of Fig. \ref{fig:1}),
and as shown on the top panel of Fig. \ref{fig:3}, a mean life of 105${^{+43}_{-28}}$ fs has been
measured.
The 969 and 1092 keV transitions depopulating this
state to the $2p_{1/2}^{-1}\otimes2^+$ symmetric one-phonon states provide branching ratios of
100(5) and 8(5), respectively, and mixing ratios, $\delta$, of $0.04(6)$ and $0.05(9)$,
respectively. Hence, the 969 keV transition to the 5/2$^-_1$ state
has a large $B(M1)$ value of
0.55${^{+0.24}_{-0.18}}$ $\mu{^2_N}$ and a small $B(E2)$ value of 0.5(2) W.u.,
while the 1092 keV transition to the 3/2$^-_1$ state exhibits a much weaker
$B(M1)$ strength of 0.03${^{+0.04}_{-0.02}}$
$\mu{^2_N}$ and
a $B(E2)$ value of 0.04${^{+0.04}_{-0.03}}$  W.u.

{\it 1840.6 keV state} ---  The  1840.6 keV level has been placed from our measurements.
The angular correlation analysis of the competing branches depopulating this state
(see Table \ref{tab:strengths} and bottom panel of Fig. \ref{fig:1})
 leads equally to  either  J$^{\pi}$=3/2$^-$ or 5/2$^-$ assignments.
A  mean life of 103$^{+35}_{-24}$ fs has been measured for this state, yielding large
$B(M1)$ values of 0.29(8) $\mu{^2_N}$ (J$^{\pi}$=3/2$^-$) and 0.28(9) $\mu{^2_N}$ (J$^{\pi}$=5/2$^-$)
for the 1153 keV transition. We proposed, however, that this state is 3/2$^-$,
based on its proximity to the 1779.7 keV level and its rather different decay strengths.

The $B(M1)$ values from the 1779.7 keV state to the 5/2$^-_1$ level 
and from the proposed 1840.6 keV state to the 3/2$^-_1$ level are greater than from any other
negative-parity states feeding the symmetric
one-phonon states. These observations, together with the appearance of these states in
the expected energy range ($\sim$ 2 MeV), support their assignment 
as the primary $MS$ states.
%Overall, as listed in Table \ref{tab:strengths}, larger $B(M1)$ values are observed in the region
%as compared with the even-A neighbors.
%Particularly remarkable is the proposed 5/2$^-$ level at 1997.6 keV level, depopulated by
%the 1186.9 keV $M1$ transition to the  5/2$^-$ one-phonon state. This transition has a large B(M1) value of
% 1.50${^{+0.21}_{-0.19}}$ $\mu{^2_N}$. However, it also  presents a very strong B(E2) contribution,
% 48${^{+26}_{-22}}$ W.u., more characteristic of a two-phonon decay.
\vspace{1mm}

\noindent {\it Interacting Boson Fermion Model} --- According to the IBM-2, even-even nuclei in the vibrational
U(5) limit exhibit $M1$ transition strengths from the  2$_{1, {\rm MS}}^+$ state
to the symmetric one-phonon  2$^+_1$ state given by,

\vspace{-0.65cm}
\begin{equation}
B(M1;2_{1, {\rm MS}}^+ \rightarrow 2^+_1)=\frac{3}{4\pi} (g_\nu - g_\pi)^2\hspace{1mm}6\hspace{1mm}
\frac{N_\nu N_\pi}{N^2} \hspace{1mm} \mu_N^2
\end{equation}

\vspace{-0.2cm}
\noindent where N$_{\pi}$ and N$_{\nu}$ are the number of proton and neutron pairs,
respectively and N=N$_{\pi}$+N$_{\nu}$. The standard boson
g-factors, g$_{\pi}$ and g$_{\nu}$, are g$_{\pi}$=1 for proton bosons
and g$_{\nu}$=0 for neutron bosons \cite{isa86}. Considering $^{88}_{38}$Sr$_{50}$ as the inert core
for the lowest $MS$ state in $^{94}$Mo \cite{3}, the proton
and neutron boson numbers are N$_{\pi}$=2 and N$_{\nu}$=1, giving
B(M1;2$^+_{{\rm MS}}$ $\rightarrow$ 2$^+_{1}$)= 0.32 $\mu{^2_N}$. 
In the weak-coupling limit, the Interacting Boson Fermion Model (IBFM) predicts that the strength of
B(M1;2$^+_{{\rm MS}}$ $\rightarrow$2$^+_{1}$) in the IBM-2 should
equal the strength of the sum of B(M1) values for  states arising from the coupling of the
unpaired particle, p, with the $MS$ state in $^{94}$Mo. Hence,

\vspace{-9mm}
\begin{center}
\begin{eqnarray}
\sum B(M1;[p\otimes 2^+_{1,MS}]_J \to[p\otimes2^+_{1}]_{3/2^-})=0.32{^{+0.12}_{-0.10}} \hspace{1mm} \mu_N^2 \nonumber \\
\sum B(M1;[p\otimes 2^+_{1,MS}]_J \to[p\otimes2^+_{1}]_{5/2^-})=0.63{^{+0.27}_{-0.21}} \hspace{1mm} \mu_N^2 \nonumber 
\end{eqnarray}
\end{center}

Whereas the former
value of 0.32$^{+0.12}_{-0.10}$ $\mu_N^2$ is consistent within the weak coupling limit, the latter value  
exceeds the schematic $U(5)$ estimate from above. This 
strong $B(M1)$ might be partly due to the spin contribution of the unpaired proton to the
$M1$ strength which is absent in the IBM-2. 

\noindent {\it Shell Model Calculations} --- To gain further insight into the nature
and structure of the proposed $MS$ states, we have examined $^{93}$Nb in the 
framework of the nuclear shell model. While a more complete description will be given
in a subsequent publication \cite{holt}, our focus here will be on providing a 
theoretical confirmation for the $MS$ interpretations proposed above and 
quantifying the spin contribution to the $M1$ transition.
%Our focus here is to provide a shell model 
%description for the $MS$ identifications proposed above and to quantify the spin 
%contribution to the $M1$ transition in this framework. 
The starting point for these 
calculations is the low-momentum nucleon-nucleon interaction $V_{low-k}$ \cite{bogner03},
% Ultimately derived from NN scattering data, $V_{low-k}$ is 
an energy independent interaction whose only free parameter, the momentum cutoff
$\Lambda$, is fixed at 2.1 fm$^{-1}$. All calculations were carried out with the
Oxbash shell model code \cite{oxbash}, which has been used with $V_{low-k}$
to reproduce the mixed-symmetry structures of both $\rm ^{92}Zr$ and 
$\rm ^{94}Mo$ \cite{holt}.
As in \cite{3} we have chosen $\rm ^{88}Sr$ as the inert core and 
used the following model space orbits for protons and neutrons: 
$\pi[2p_{1/2}$, $1g_{9/2}$, $1g_{7/2}$, $2d_{5/2}$, $2d_{3/2}$, $3s_{1/2}]$ and 
$\nu[1g_{7/2}$, $2d_{5/2}$, $2d_{3/2}$, $3s_{1/2}$, $1h_{11/2}]$, where the single-particle
energies were optimized to fit the experimental spectra
of $\rm ^{90}Zr$ and $\rm ^{90}Sr$. To make our current calculations fully
predictive, we fixed the effective quadrupole charges, $e_p=1.85e$ and 
$e_n=1.3e$, as the averages of the values used for $^{92}$Zr and $^{94}$Mo 
and quenched the spin $g$-factors by 0.7.

The relevant SM results are shown in Table I, where we include theoretical
$B(M1)$ and $B(E2)$ values with the isoscalar and isovector components of the $E2$ 
transition from the 1/2$^-$ level.
%, where
%\begin{eqnarray}
%IS=<Q_p>+<Q_n> \\ \nonumber
%IV=<Q_p>-<Q_n>.
%\end{eqnarray}
We list only those SM states 
that have a clear manifestation in the experimental data, identifying the 1396 keV
$7/2^-$ and 1500 keV $9/2^-$ states as the splitting of the $4^+_1$ two-phonon 
state and the 1572 keV $3/2^-$ and 1588 keV $5/2^-$ states as the splitting of the 
two-phonon $2^+_2$ state.

% Of the low-lying experimental levels, only those at 1285
%keV and 1370 keV were not obviously reproduced in the SM calculations.
%where there is otherwise generally good agreement between theory and experiment.

%The only discrepancy being a reversed assignment of MS and two-phonon character for
%the $2^+_2$ and $2^+_3$ in $\rm ^{94}Mo$.

%and understanding the large $B(M1)$ 
%values found for the $5/2^-_{\rm MS}$ state, we .

%Traditionally experimental $B(E2)$ and $B(M1)$ measurements were used to 
%confidently identify MS states, and shell model calculations were performed after
%the fact to describe the experimental observations. In odd-mass nuclei the unpaired
% proton can potentially complicate the picture by contributing a misleading 

%It can thus be used in any nuclear region with no alterations, and with 
%thederived effective charges and $g$-factors discussed below, the results presented
%represent true predictions.  As in \cite{holt} 

{\it 1779.7 keV $5/2^-$ state} --- We identify the $5/2^-_2$ SM
state (at an energy of 1734 keV) with the experimentally observed 1779.7 keV
state. In Table I we see the SM predicts $B(M1)$ values to the fully symmetric one-phonon
$5/2^-$ and $3/2^-$ 
states that are in good agreement with the experimental results, noting in particular the 
strong $M1$ transition to the $5/2^-$ final state. By analyzing the dominant components of 
the calculated SM wavefunctions \cite{holt}, we can verify
that the $^{93}$Nb states result from the orthogonal bosonic core 
states of $^{94}$Mo coupling to a $p_{1/2}$ hole. Under this assumption,
the ratio $B(M1: 5/2^-_{MS} \rightarrow 5/2^-_1) / B(M1: 5/2^-_{MS} 
\rightarrow 3/2^-_1)$ can be expressed as a ratio of Racah coefficients, from which
%The significant suppression in $M1$ strength for the transition to the $J-1$ 
%state is a result of the reduced $M1$ matrix elements of
%$\frac{5}{2}^-_{ms} \rightarrow \frac{5}{2}^-_1,\frac{3}{2}^-_1 $ being related
%through the factors $\sqrt{2J_f+1}$ $ \left\{ {2 \atop 5/2}{J_f\atop 2}{1/2\atop 1} \right\}$,
%where $J_f$ denotes the angular momentum of the final states.  
we predict the $B(M1)$ value for the $J_i \neq J_f$ transition
be suppressed by a factor of 14 with respect to the $J_i=J_f$ transition. This is 
clearly seen in both the calculations and experiment.
%likely arising from the additional spin-flip amplitude involved.  jdh
%Likewise, the small $B(E2)$ values to the same states agree with experiment.
A final theoretical prediction consistent with the $MS$ identification of this state is the 
weakly-collective $E2$ transition with significant isovector character
to the $1/2^-$ first excited state.

We now examine the spin and orbital parts of the $B(M1)$ strength to see if
the shell model predicts an enhanced spin contribution from the unpaired proton.
Using vanishing orbital $g$-factors ($g^l_p=g^l_n=0$), we find the spin $B(M1)$ 
value to be 0.155 $\mu^2_N$, and with vanishing spin $g$-factors ($g^s_p=g^s_n=0$), 
we have the orbital $B(M1)$ value of 0.153 $\mu^2_N$. Since a similar finding
was reported for $MS$ states in even-mass nuclei \cite{3}, it
appears that there is little significant spin enhancement to the $B(M1)$ 
values.

{\it 1840.6 keV state} --- The shell model predicts a 1811 keV $3/2^-_{\rm MS}$
spin-flip partner to the $5/2^-_{\rm MS}$ state. An obvious candidate would be the 1840.6 keV
experimental state. In Table I we see that, however, there
are two possible $J$ assignments for this level. Theoretical 
considerations support the labeling of this state as  $3/2^-_{\rm MS}$. Primarily,
its close proximity in energy to the 1779.7 keV $5/2^-$ state suggests its candidacy as 
the spin-flip partner. It also exhibits a strong $M1$ transition to 
the symmetric $3/2^-$ one-phonon state and a suppressed (now by a theoretically-predicted
factor of 9) $M1$ 
to the symmetric  $5/2^-$  one-phonon state,
%Since we expect a strong $M1$ to the same $J$ and a suppressed $M1$ to $J+1$, this
%would 
further confirming the $3/2^-_{MS}$ assignment. With a $3/2^-$ label for this state, 
the shell model predicts a weakly-collective
$E2$ transition with significant isovector character to the $1/2^-$ first excited state,
just as with the $5/2^-_{\rm MS}$ state identified above.
Similarly, we see no sign of an enhanced spin contribution to the
$M1$ transition, where $B(M1)_{s}=0.110$ $\mu^2_N$ and $B(M1)_{l}=0.121$ $\mu^2_N$, and
conclude that this state can reasonably be identified as $3/2^-_{\rm MS}$. 
In conclusion, we propose, for the first time, $MS$ states in a nearly spherical
odd-mass nucleus from both experimental and theoretical evidence.

%one-phonon states, we see that this state exhibits a strong $M1$ transition to the
%one phonon $3/2^-$ state and a suppressed $M1$ to the one-phonon $5/2^-$ state.
%This is in strong agreement with the shell model results for a $3/2^-$ initial stat
%where we would expect a strong $M1$ transition to the one-phonon state with similar
%$J$, but a suppressed $M1$ to the state with $J+1$, as discussed above.  
%Furthermore, with 

%From the large B(E2) transitions to the $\frac{1}{2}^-_1$ ground state, the 
%$\frac{3}{2}^-_1$ and $\frac{3}{2}^-_1$ are clearly identified as the splittings of
%the $2^+_1$ one-phonon state of $\rm ^{92}Zr$ and $\rm ^{94}Mo$.

%Finally, the corresponding $3/2^-_{\rm MS}$ and the quintuplet of MS states associated with the
%(2$_{1,{\rm MS}}^{+}$)$\otimes$ ( 1$g_{9/2}$, $^{92}$Zr) particle-core coupling has not yet
%been identified.

%\newpage

\vspace{-0mm}
We would like to thank H.E. Baber for technical support and
B.A. Brown for his  assistance with the Oxbash code.
This work was supported by the U.S. National Science Foundation under Grants No. PHY-0354656
and PHY-0245018, and by the German DFG under Grants No. Pi 393/1-2 and Jo 391/3-1.

% BibTeX users please use
% \bibliographystyle{}
% \bibliography{}

\end{document}